\documentclass[]{spie}  
\usepackage[dvipdfmx]{graphicx}
\usepackage[varg]{txfonts}

\title{DIOS: the dark baryon exploring mission}

\author{T.~Ohashi\supit{a}, Y.~Ishisaki\supit{a}, Y.~Ezoe\supit{a}, S.~Yamada\supit{a},
S.~Yamaguchi\supit{a}, N.~Miyazaki\supit{a},
Y.~Tawara\supit{b}, K.~Mitsuda\supit{c},
  N.~Y.~Yamasaki\supit{c}, Y.~Takei\supit{c}, K.~Sakai\supit{c}, K.~Nagayoshi\supit{c}, R.~Yamamoto\supit{c}, A.~Chiba\supit{c}, and T.~Hayashi\supit{c}
\skiplinehalf
  \supit{a} Department of Physics, Tokyo Metropolitan University, 1-1 Minami-Osawa, Hachioji, Tokyo~192-0397, Japan\\
  \supit{b} Depertment of Phycics, Nagoya University, Furo-cho, Chikusa, Nagoya~464-8602, Japan\\
  \supit{c} Institute of Space and Astronautical Science, Japan Aerospace Exploration Agency, 3-1-1, Yoshinodai, Chuo, Sagamihara, Kanagawa~252-5210, Japan\\
}

\authorinfo{Further author information: (Send correspondence to T. Ohashi.)\\T. Ohashi: E-mail: ohashi@tmu.ac.jp, Telephone: +81-426-77-2492}

\begin{document} 
\maketitle 

\begin{abstract}
DIOS (Diffuse Intergalactic Oxygen Surveyor) is a small satellite
aiming for a launch around 2020 with JAXA's Epsilon rocket. Its main
aim is a search for warm-hot intergalactic medium with high-resolution
X-ray spectroscopy of redshifted emission lines from OVII and OVIII
ions. The superior energy resolution of TES microcalorimeters combined
with a very wide field of view (30--50 arcmin diameter) will enable us
to look into gas dynamics of cosmic plasmas in a wide range of spatial
scales from Earth's magnetosphere to unvirialized regions of clusters
of galaxies. Mechanical and thermal design of the spacecraft and
development of the TES calorimeter system are described. We also
consider revising the payload design to optimize the scientific
capability allowed by the boundary conditions of the small mission.

\end{abstract}



\section{INTRODUCTION}
\label{sect:intro}  

Majority of baryons in the local universe remain unexplored and are
called as dark baryons. Baryon census has been carried out based on
the observed contents of widespread medium around galaxies, clusters,
and in large-scale structures.  However, 30--50\% of the baryons are
still missing\cite{shull12}. Numerical simulations and recent
observations indicate that dark baryons are likely to reside along the
large-scale filaments in the form of warm-hot intergalactic medium
(WHIM) with temperatures between $10^5$ and $10^7$ K\@.  The density
of WHIM is 10--100 times that of the average level in the universe,
and even lower than those in the virial radii of clusters of galaxies
by a factor of up to 10. This makes the direct detection of WHIM very
difficult. Note that UV observations of OVI and other absorption lines
from FUSE and HST (COS in particular) have given firm evidences that
there indeed exists a warm gas with temperatures $\sim 10^5$ K in the
intergalactic space\cite{shull12}.

Since main part of WHIM is in a temperature range of a few times
$10^6$ K, the most efficient observation will be in X-rays, either by
emission or absorption lines. There are a few cases where absorption
by the WHIM is observed significantly with grating spectrometers on
Chandra and XMM-Newton\cite{nicastro13,ren14}.  As for the WHIM
detection in emission, the low surface brightness makes the study
particularly difficult. However, X-ray emission from WHIM will tell us
about its spatial distribution at different redshifts and will show us
the evolution of the large-scale structure of the universe in a direct
way.

The concept and design of DIOS (Diffuse Intergalactic Oxygen Surveyor)
has been reported in the past SPIE conference papers for several
times\cite{ohashi06,tawara08,ohashi10,ohashi12}.  A special feature of
the DIOS mission is a wide-field X-ray spectroscopy using an array of
microcalorimeters, enabling the detection of WHIM in emission. A
detailed simulation has been carried out assuming a few times larger
X-ray telescope\cite{takei11}, and the results can be easily scaled to
DIOS observations since influence of non-X-ray background will be
negligible.  Technology of X-ray microcalorimeters has been well
established toward the launch of ASTRO-H in late
2015\cite{takahashi12}. This mission will be the first satellite to
carry out cosmic X-ray observations with microcalorimeters. New
technologies which enable the space application of microcalorimeters
have been developed: such as Joule-Thomson coolers backing up liquid
He cooling, 3-stage ADR (adiabatic demagnetization refrigerator)
maintaining the detector temperature at 50 mK, an on-board pulse shape
analysis system, and X-ray generators enabling continuous calibration
in the orbit.

The unique capability of microcalorimeters is the high-resolution
spectroscopy of extended sources, in contrast to grating
spectrometers. ASTRO-H SXS experiment gives a field of view of $3
\times 3$ arcmin, and studies of much more extended objects will have
to await further development of, such as, a TES calorimeter array.
This type of instrument is a baseline instrument for Athena, the large
X-ray observatory planned for launch in 2028 by ESA\@.  DIOS will
provide a large field of view (about 50 arcmin) by combining TES
calorimeters with a short focal length telescope. X-ray spectroscopy
of extended objects from earth's environment to cluster outskirts will
be carried out in a very efficient way. Thermalization processes,
shock formation, turbulence, resonance scattering and charge exchange
process will all be examined in a clear way.

In this paper, we give an updated status of the DIOS mission over the
previous reports.  We plan to propose DIOS to the 4th mission in the
JAXA series of small project using Epsilon rocket for the
launch. JAXA's framework for small missions has been revised in 2013
with the release of a new roadmap of space science and space
exploration in Japan. We will discuss possible ways of enhancing the
original capability of DIOS within the available payload
specifications. We note that DIOS is along the line of continuous
effort of an international collaboration which so far has proposed
dark-baryon missions to ESA's Cosmic Vision (EDGE and ORIGIN) and to
the US Decadal Survey (Xenia)
\cite{denherder07,burrows10,denherder11,denherder12}.

\begin{figure}[!htb] 
\begin{minipage}{0.48\textwidth}
\includegraphics[width=0.95\textwidth,bb=0 0 722 450]{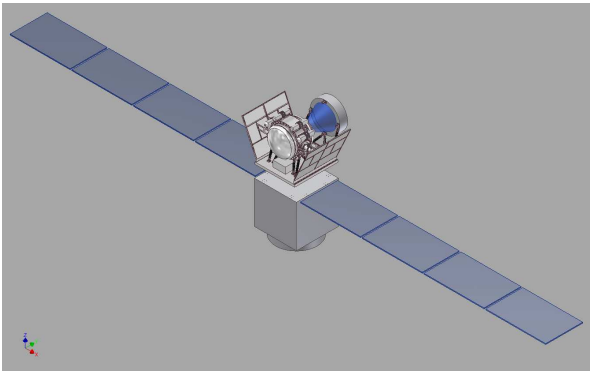}
\caption{The DIOS spacecraft. The length of the solar paddle is 10 m after it is deployed. Total mass will be about 615 kg.}
\label{spacecraft}
\end{minipage}
\newcounter{keepfignum}
\setcounter{keepfignum}{\value{figure}}
\renewcommand{\figurename}{Table}
\setcounter{figure}{\value{table}}
\addtocounter{table}{1}
\begin{minipage}{0.5\textwidth}
\caption{Parameters of the DIOS spacecraft}
\begin{center}
\begin{tabular}{|l|l|}\hline
Total mass & 615 kg \\
Payload mass & 323 kg \\
Size at launch & $1.2 \times 1.45\times 1.4$ m\\
Size in orbit & $5.9 \times 1.45 \times 1.4$ m \\
Attitude control & 3-axis \\
Pointing accuracy & $\le 30$ arcsec \\
Total power & 691 W \\ 
Payload power & 381 W \\ \hline
\end{tabular} \end{center}
\end{minipage}
\end{figure}
\renewcommand{\figurename}{Figure}
\setcounter{figure}{\value{keepfignum}}

\begin{figure}[!htb] \begin{minipage}{8cm}
\includegraphics[width=0.9\textwidth,bb=0 0 842 504]{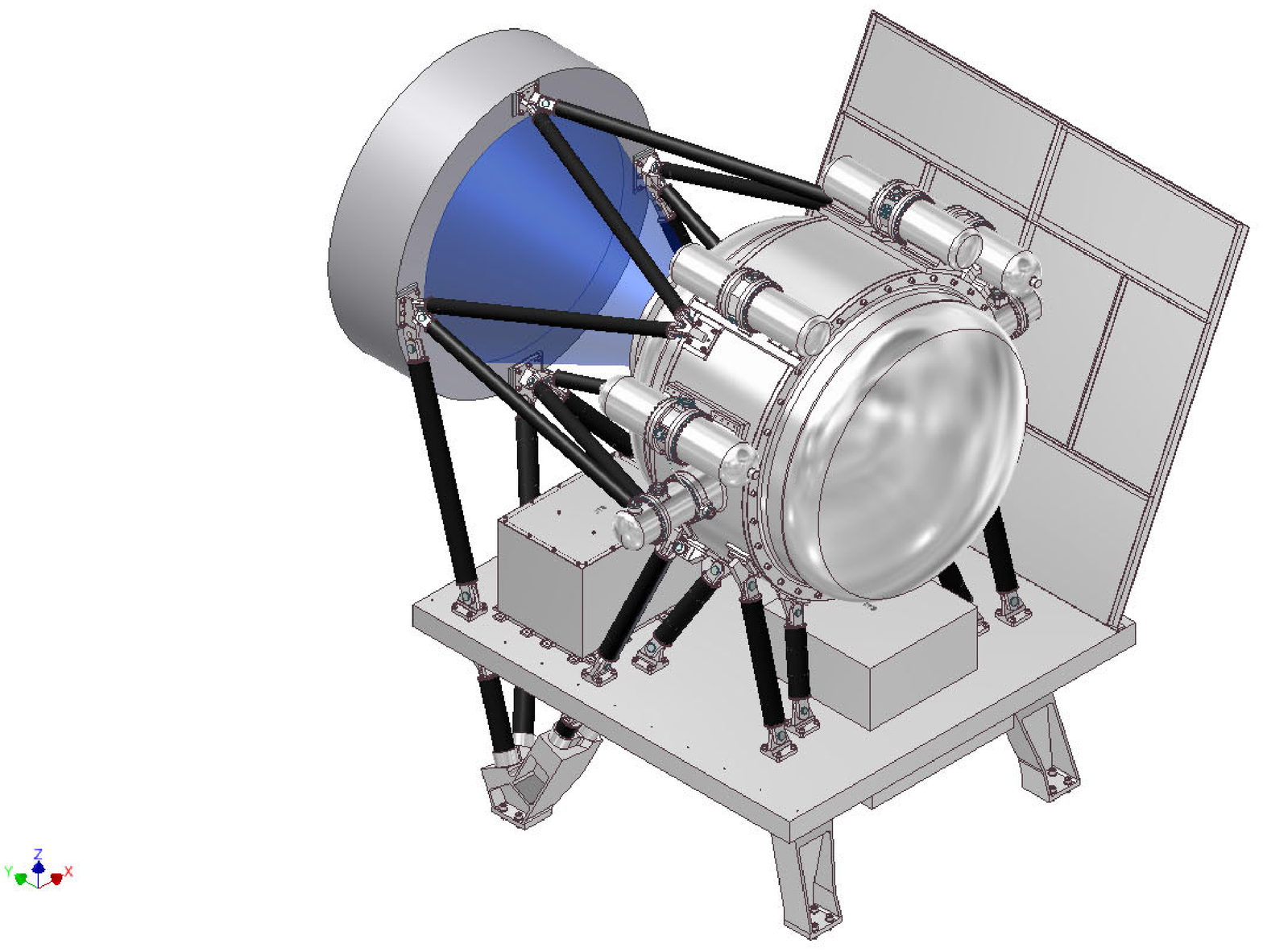}
\caption{Close-up view of the payload of DIOS, showing the X-ray
  telescope and dewar. The focal length of the telescope is 70
  cm.} \label{dios_view2}
\end{minipage} \hfill
\begin{minipage}{8cm}
\includegraphics[width=0.85\textwidth,bb=0 0 191 112]{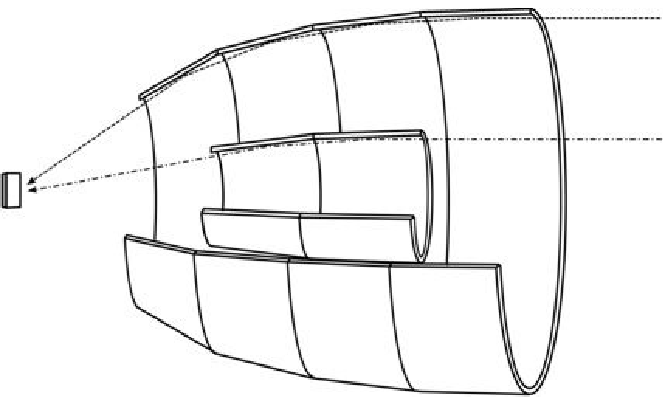}
\caption{Concept of the four-reflection X-ray telescope for DIOS.}
\label{FXT-concept}
\end{minipage}\end{figure}

\section{Spacecraft} 

The view of the DIOS spacecraft is shown in Fig.\ \ref{spacecraft},
and main parameters of the satellite are summarized in Table 1. The
spacecraft will weigh about 600 kg, with the payload mass $\sim 320$
kg.  The total spacecraft power will be $\sim 700$ W, of which about
380 W will be used by the payload. The cryocoolers (2-stage Stirling
coolers and Joule-Thomson coolers) will require power around 300 W\@.
The orbit of DIOS will be a low-earth circular one with an altitude
of about 550 km, same as those of all the previous Japanese X-ray
satellites including Suzaku.  The launch will take place at USC
(Uchinoura Space Center, $131.1^\circ{\rm E}, 31.1^\circ{\rm N}$) in
Kagoshima prefecture, Japan, giving the orbital inclination to be
$31^\circ$.

The spacecraft attitude will be 3-axis stabilized. Since angular
resolution of the X-ray instrument will be about $3'$, DIOS will not
need very fine pointing accuracy. To reconstruct the attitude
accurately enough for the data analysis, star trackers will be equipped
and give the attitude information with about $10''$ accuracy. The
baseline design has the telescope direction perpendicular to the sun
direction, and accessible sky region will be along a great circle with
$90^\circ \pm 25^\circ$ from the sun direction. Radiator panels with an
approximate area of 1 m$^2$ will exhaust heat from the spacecraft. Thermal
analysis indicates that heat input when the radiator points to the
earth will not be a serious problem, so switching of two radiator
panels is not an absolute necessity.

\begin{figure}[!htb] 
\begin{minipage}{8cm}
\includegraphics[width=0.9\textwidth,bb=0 0 234 258]{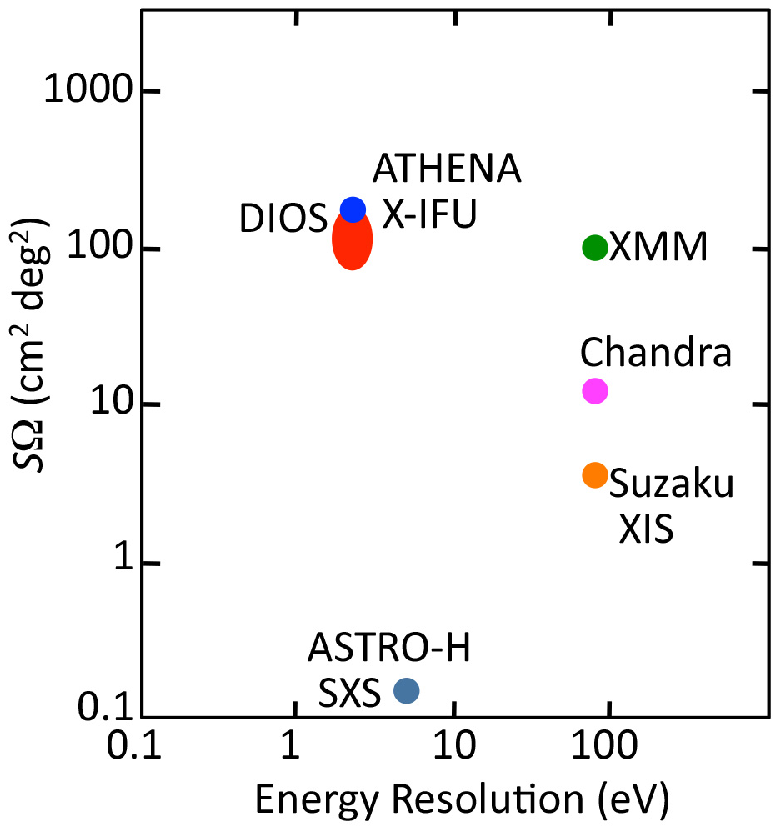}
\caption{Comparison of $S\Omega$ and energy resolution for
spectroscopic instruments (CCDs and microcalorimeters) for planned and
operating X-ray satellites.} \label{hikaku}
\end{minipage}
\newcounter{keepfignum1}
\setcounter{keepfignum1}{\value{figure}}
\renewcommand{\figurename}{Table}
\setcounter{figure}{\value{table}}
\addtocounter{table}{1}
\begin{minipage}{8cm}
\caption{Parameters of the observing instruments on board DIOS}
\label{instr_param}
\begin{center}
\begin{tabular}{|l|l|}\hline
Effective area & $200$ cm$^2$ \\
Field of view & $50'$ diameter \\
$S\Omega$ & $\sim 150$ cm$^2$ deg$^2$ \\
Angular resolution & $3'\ (16\times 16$ pixels) \\
Energy resolution & $< 5$ eV (FWHM) \\
Energy range & 0.1-- 1.5 keV \\
Observing life & $> 5$ yr \\ \hline
\end{tabular} \end{center}
\end{minipage}
\end{figure}
\renewcommand{\figurename}{Figure}
\setcounter{figure}{\value{keepfignum1}}

A close-up view of the payload part is shown in
Fig.\ \ref{dios_view2}. The payload consists of a four reflection
telescope (FXT whose design concept is shown in
Fig.\ \ref{FXT-concept}) and a dewar containing X-ray spectrometer
array (XSA) consisting of TES microcalorimeters. Focal length of FXT
is 70 cm in the baseline design. FXT and dewar are connected and
supported from the baseplate by CFRP torus structures. The baseplate
also provides an interface between the payload part and the satellite
bus. The baseplate will be equipped with electronics boxes and a
radiator panel will stand out with an angle not directly looking into
solar paddles.  The dewar is similar to but smaller than the ASTRO-H
one since no liquid He will be used in DIOS\@. Thermal and mechanical
analysis has been carried out for a simplified model. We confirmed
that cooler powers already confirmed for ASTRO-H will be high enough
to cool the DIOS instrument with an input power of 280 W\@. Also, the
characteristic frequency of the payload is within the acceptable
range. We plan to launch the spacecraft in warm condition, and to
spend a few weeks to cool down the payload in space. In order to
generate the total power of about 700 W, the solar paddle consists of
4 panels on both sides.

Fig.\ \ref{hikaku} shows $S\Omega$ or grasp for various missions
against energy resolution, and Table \ref{instr_param} shows
parameters of the DIOS instrument. Energy resolution needs to be
better than 5 eV to resolve red-shifted WHIM emission lines from those
of Galactic or solar-system emission.  DIOS will provide a factor of
$\sim 250$ improvement over ASTRO-H SXS in $S\Omega$, which is close
to the level of Athena X-IFU\@. With such a small satellite, DIOS will
offer a very high sensitivity to extended X-ray emission.

\section{Instrument development}

\subsection{X-ray telescope}
Development of FXT (4-reflection X-ray telescope) in Nagoya University
was started early and has been reported for a number of
times\cite{tawara04,tawara06,tawara10} as well as in the present
conference (9144-235). A schematic view of FXT is shown in Fig.\
\ref{FXT-concept}. We summarize its performance briefly. The focal
length is 70 cm, enabling a small focal-plane instrument to cover a
large sky area and leading to very low detector background. The solid
angle of the field of view of DIOS ($50'$) will be about 300 times
larger than the SXS instrument on ASTRO-H\@. The goal of the angular
resolution is $3'$, noting that four reflections inevitably introduce
additional image blurring than the 2 reflection mirror. An X-ray beam
measurement of a test mirror set in a quadrant housing gave a
half-power diameter of about 10 arcmin, and use of thicker foils (0.2
mm rather than 0.15 mm) is under consideration.

\begin{figure}[!htb] 
\begin{minipage}{7cm}
\centerline{\includegraphics[width=0.9\textwidth,bb=0 0 295 199]{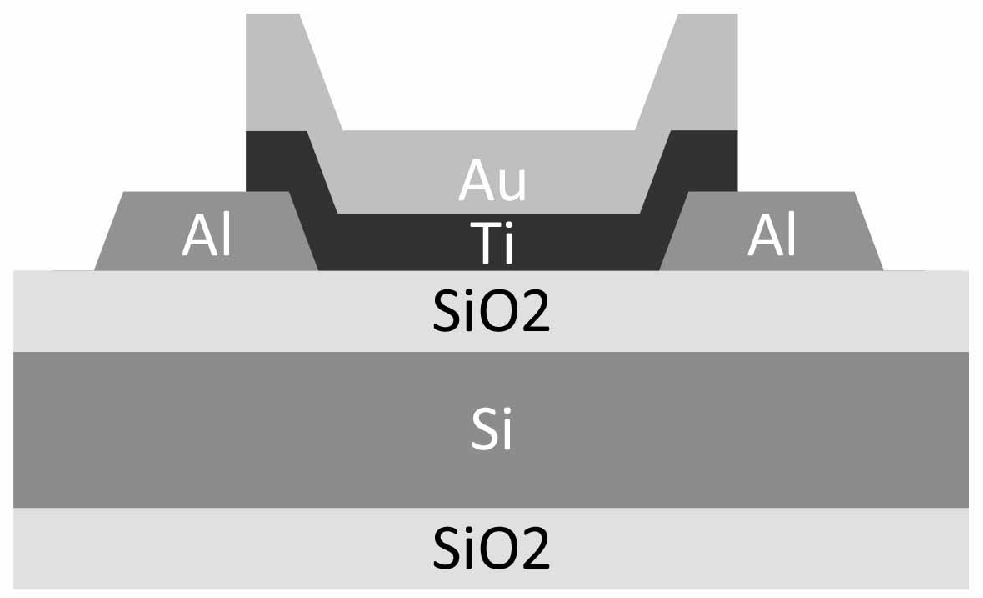}}
\caption{Cross sectional view of a TES pixel for the new layered
  wiring developed for TES array.} \label{sekisou3}
\end{minipage} \hfill
\begin{minipage}{9cm}
\centerline{\includegraphics[width=0.9\textwidth,bb=0 0 251 196]{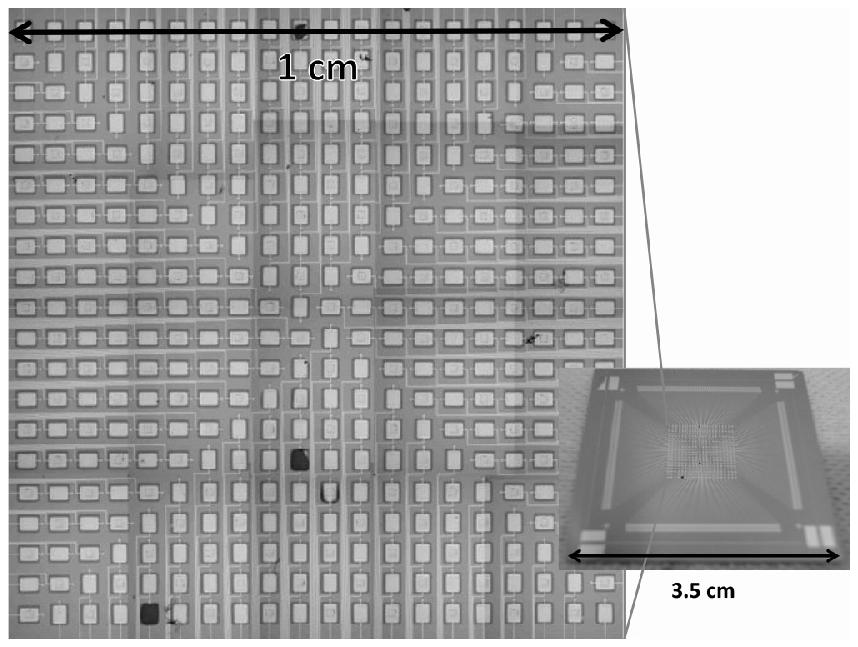}}
\caption{A test model of TES calorimeter array with $20 \times 20$
  pixels.} \label{400TES}
\end{minipage}
\end{figure}

\subsection{TES array}
The focal plane instrument XSA (X-ray Spectrometer Array) is an array
of TES microcalorimeters, whose development in Japan is a
collaborative effort of Tokyo Metropolitan University, ISAS/JAXA,
and Kanazawa University.  Below, we briefly describe the status of our
instrument development.

Since the wirings all running in the same plane will hamper dense
layout of the TES calorimeter pixels, we have been developing
multilayer wiring technique \cite{yamada14}. A cross section of a TES
pixel is shown in Fig.\ \ref{sekisou3}, and a test array with 400
pixels employing the multi-layer wiring is shown in Fig.\ \ref{400TES}.
The hot and return lines run in different layers separated by an
insulation layer. These wires make contact at a contact hole near each
calorimeter pixel.  The Si substrate has a thickness of 300
$\mu$m. The wirings are made of Al, and their thicknesses are 50--100
nm, and the insulation SiO$_2$ layer has a thickness of 500 nm. Width
of the wirings is only 10 and 15 $\mu$m, and a very fine alignment is
performed during the production. Our production process requires that
the multi-layer wirings are built first in the Si substrate, and then
we form TES calorimeters on top of these wirings.

Early practice showed some problems which required modification in the 
process of
the array production\cite{oishi12}. The wiring makes a step of about
50 nm height at the edge of the square pixel area, and simple
deposition of Ti-Au bilayer causes damage of the bilayer structure due
to a mechanical break at these steps. Because of this, we could not
obtain a good superconducting transition performance, such as a dull
$R-T$ curve.

To improve this, we introduced a slanted machining of the wall of the
multi-layer wiring and created a slope of about $45^\circ$ as shown in
Fig.\ \ref{sekisou3}.  This additional machining is provided by
National Institute of Advanced Industrial Science and Technology
(AIST) in Japan. We find that forming of Ti-Au bilayer on this slope
does not cause a mechanical break, and the quality of the bilayer is
improved.  A test sample with 40 nm thick Ti and 130 nm Au bilayer
showed a significantly better transition properties with a transition
temperature 175 mK, residual resistivity of 0.3 m$\Omega$, and normal
resistivity of 378 m$\Omega$. Therefore, basic production method of
TES array with multilayer wiring has been almost established. Next
step is a production of actual TES array and its test with X-rays.

We performed an experiment to examine radiation tolerance of TES
calorimeters\cite{ishisaki14}.  TES calorimeter with a Ti/Au bilayer
of 30/40 nm thick attached with a 1.5 $\mu$m thick Au absorber was
irradiated by 150 MeV proton beam with a total dose of 10 krad. This
is approximately 10 years of irradiation in the low Earth orbit. We
see no significant change in the transition temperature or the energy
resolution, which is $5.6\pm 0.3$ eV (FWHM) at 6 keV after the
irradiation and degraded by about 10\%.  This indicates that our TES
calorimeters have sufficient radiation tolerance in orbit.

\subsection{TES array readout}
Frequency Domain Multiplexing (FDM) method to handle the signal
readout from TES arrays has been developed\cite{ohashi12}. TES pixels are
AC-biased with different frequencies at an order of MHz, and summed by
a SQUID and de-multiplexed by room temperature electronics.  We
designed, fabricated and tested several key components.

Phase delay between the SQUID and the room temperature demultiplexing
electronics limits the bandwidth of the SQUID to be less than $< 1$
MHz, and it affects the number of multiplexing channels with a
standard flux-locked loop feedback.  We adopt a base-band feedback
(BBFB) method\cite{takei09} to compensate the delay by adjusting the
phase in the feedback loop.  An analog electronics using phase
sensitive detectors are developed and demonstrated for a 4-channel TES
device at bias frequencies of 1 -- 3 MHz.  We successfully detected
X-ray pulses simultaneously from the TES array, but deficiency of the
loop-gain caused non-linearity in SQUID and produced crosstalks
between TES pixels\cite{yamamoto14}.

\begin{figure}[!htb]
\begin{center}
\includegraphics[bb=0 0 730 202,width=0.85\textwidth]{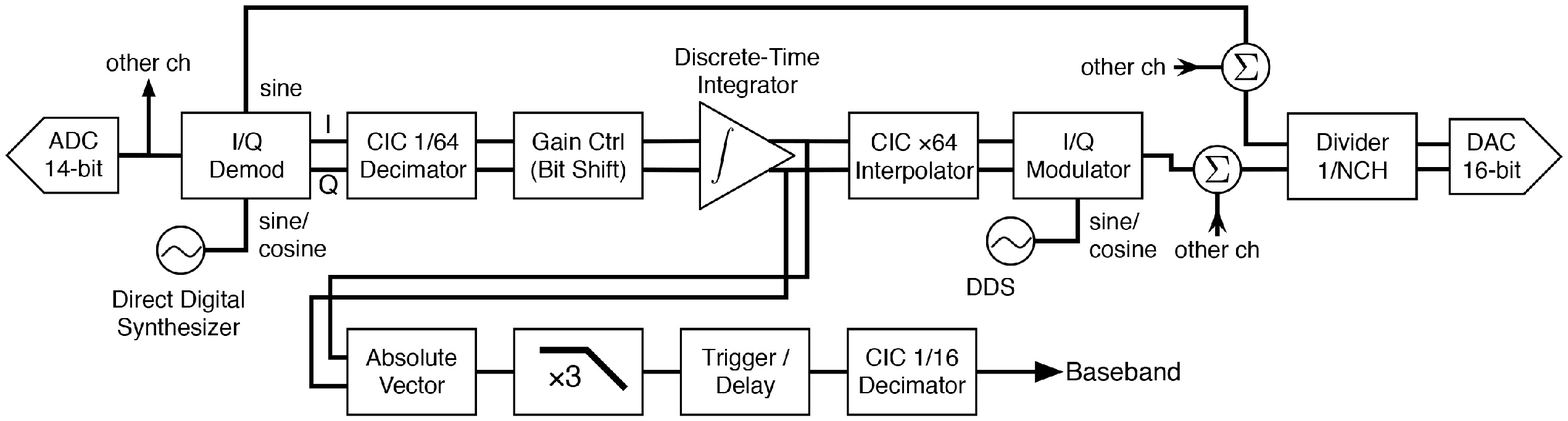}
\includegraphics[bb=0 0 432 288,width=0.4\textwidth]{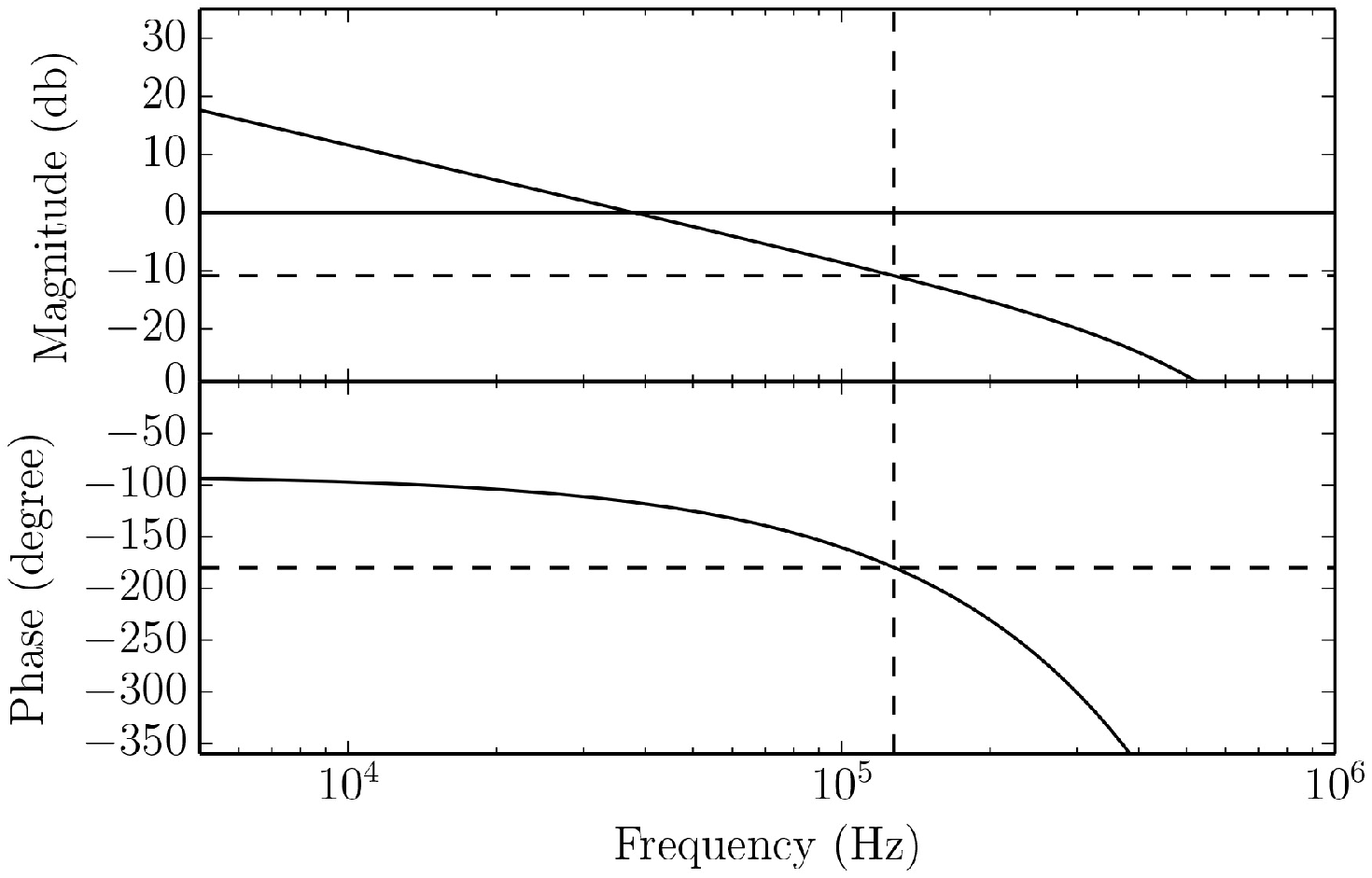}
\hspace{1cm}
\includegraphics[bb=0 0 307 237,width=0.4\textwidth]{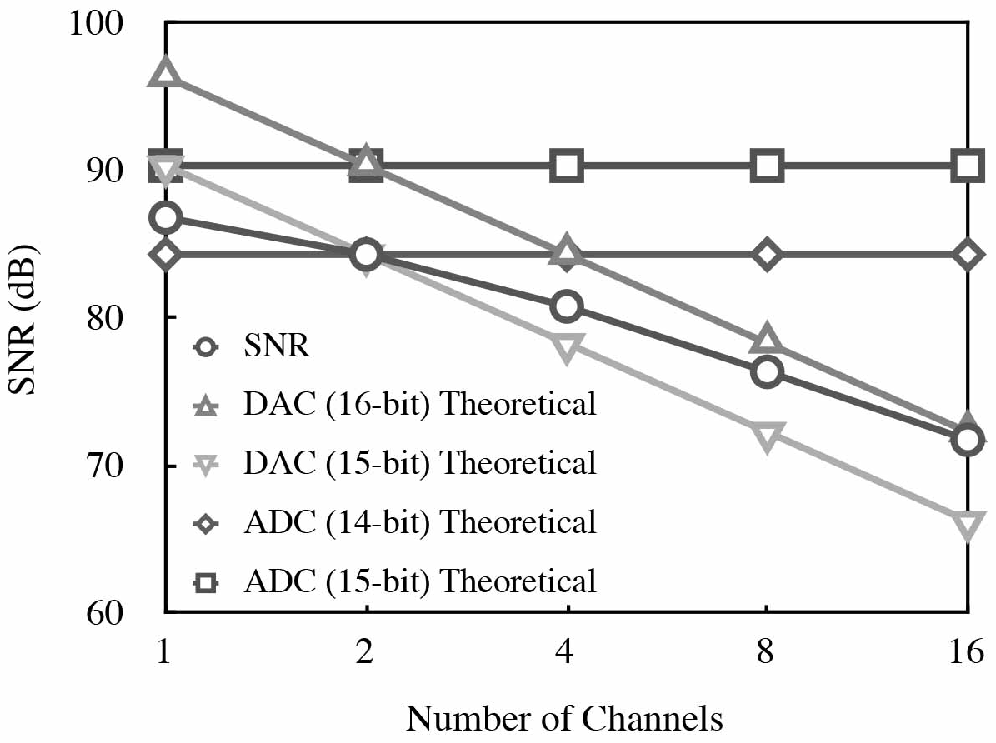}
\caption{(Top) Current design of BBFB circuit using FPGA, 
(Bottom-left) Loop-gain and phase of current design. 
(Bottom-right) Signal-to-Noise ratio as a function of number of multiplexed channels with 
DAC/ADC resolution.}
\label{fig:Digitals}
\end{center}
\end{figure}

Based on these experiments, we designed and manufactured a digital
BBFB circuit using FPGA in order to demultiplex and trigger
events\cite{sakai14} as shown in Fig.\ \ref{fig:Digitals}.  The
loop-gain was improved to handle driving frequencies up to 7 MHz.  The
number of multiplex channels is limited by the bandwidth and
resolution of DAC in the output feedback current.  We tested the
performance against simulated TES pulses for 16 channels with a 14-bit
250 Msps (Mega-samples per sec) ADC and a 16-bit 800 Msps DAC, and obtained
enough signal-to-noise ratio of 72 dB where 60 dB was required to
achieve a 2eV resolution for pulses corresponding to 1 keV\@.  The
expected resolution is also shown in Fig.\ \ref{fig:Digitals}. For
DIOS detector system, a 15 bit DAC is required to multiplex 16 channels
by keeping sufficient energy resolution.
  
\begin{figure}[!htb]
\begin{center} 
\includegraphics[width=0.3\textwidth,bb=0 0 1340 1340]{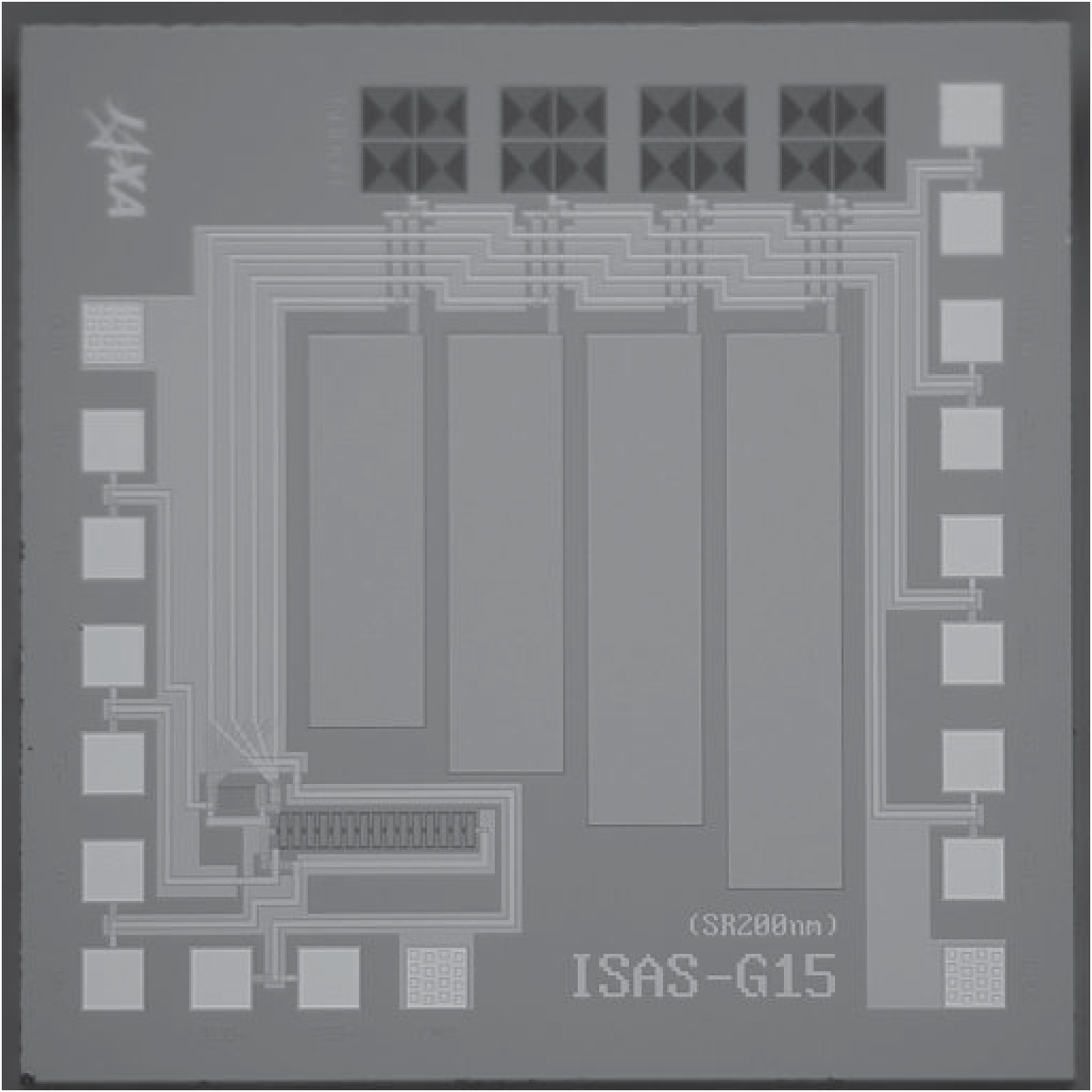}
\hspace{2cm}
\includegraphics[width=0.4\textwidth,bb=0 0 768 528]{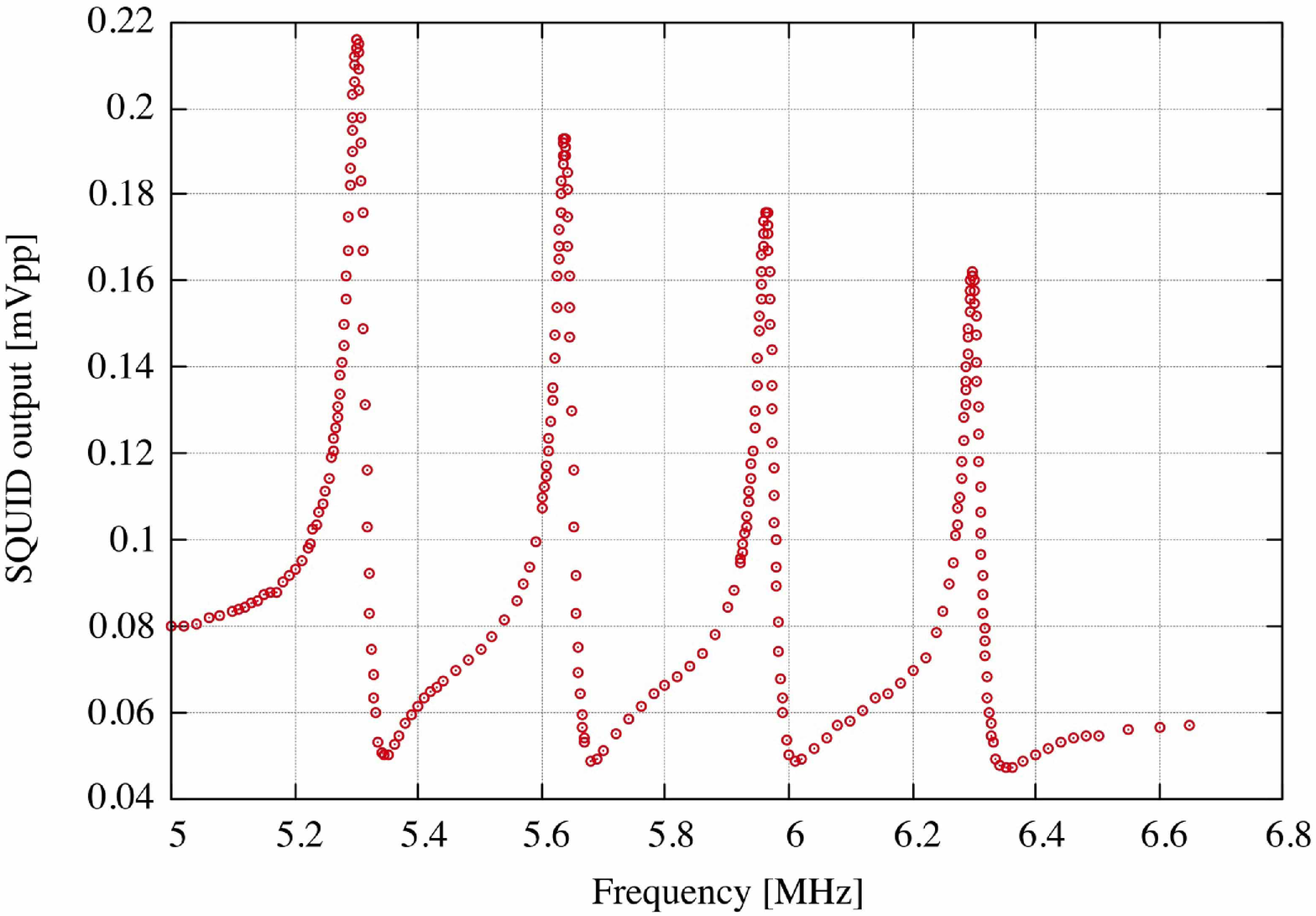}
\caption{(Left) A picture of LC resonators on a 2.5$\times$2.5 mm chip. 4 capacitors of 
different area and 4 500nH coils are fabricated with a SQUID array. 
(Right) Measured response of LC resonators at He temperature.}
\label{fig:Filter}
\end{center}
\end{figure}

Suppression of power dissipation by SQUID is important, because it
limits configuration of cold stage components and distance between
SQUID and TES array. We developed SQUID series for this application
whose power dissipation is as low as $\sim 30$ nW per channel with
enough gain\cite{sakai14}. Gradiometer at SQUID input-coil works
effectively under the Earth's magnetism.  LC-filter array is also
fabricated on $2.5 \times 2.5$ mm$^{2}$ SQUID chips as shown in Fig.\
\ref{fig:Filter}.  4 capacitors with different area are made with
anodized Aluminum.  In combination with 500 nH coils, resonances
between 5 and 6 MHz are measured by a SQUID at the He temperature.
End-to-end test which includes a TES array, on-chip resonator,
low-power SQUID, and digital BBFB circuit are planned.

\begin{figure}[!htb] \begin{center}
\centerline{\includegraphics[bb=0 0 462 251,width=0.7\textwidth]{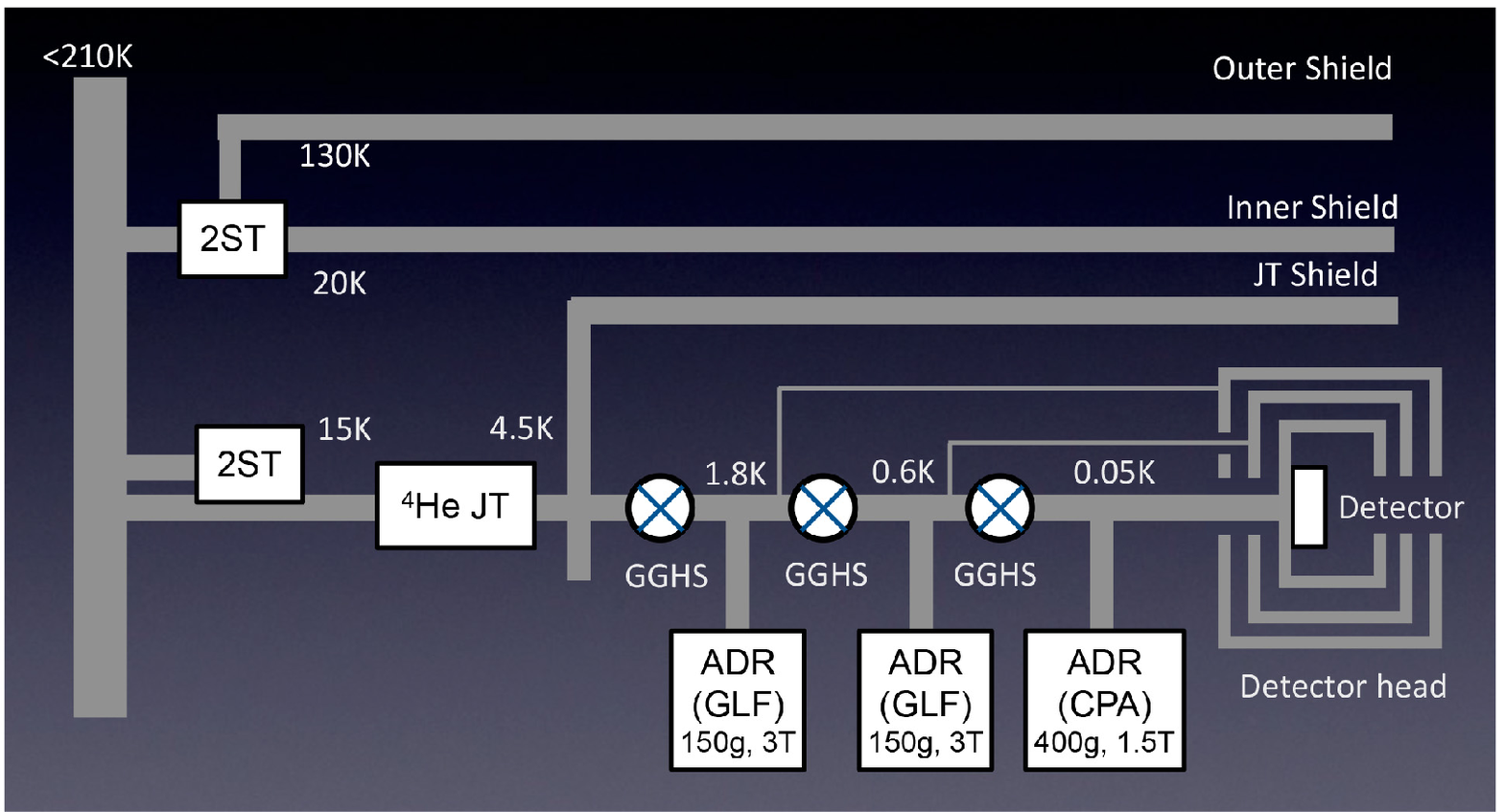}}
\caption{Cooling chain of DIOS, consisting of 2-stage Stirling coolers, $^4$He JT cooler and 3 ADRs.} \label{coolingchain}
\end{center} \end{figure}

\section{Thermal and mechanical design}


The cooling system of DIOS is the same as that employed for
ASTRO-H\cite{ysato12}, except that no liquid He will be used for
DIOS\@.  The cooling chain is shown in Fig.\ \ref{coolingchain}.  We
will keep the radiator panel with an area of 1.5 m$^2$ at less than
210 K\@.  The first stage of the cooler is a 2-stage Stirling cooler
which takes the temperature down to 15--20 K\@.  Then, $^4$He
Joule-Thomson cooler takes the temperature down to 4.5 K\@. Cool-down
tests have been performed for both engineering (EM) and flight (FM)
models of the ASTRO-H dewar, and coolings without employing liquid He
show good performances.  No consumable cryogen will be used in the
DIOS cooling system, and we expect to have a long observing life in
the orbit if mechanical coolers continue to work. The DIOS coolers
will incorporate all the experimental knowledge obtain in the test of
ASTRO-H coolers. In particular, micro-vibration influence to the TES
detectors has been addressed and further reduction of this effect will
be the subject of cooler development for DIOS\@.  The XSA system will
go through a warm launch, and we have to allow for the initial
cooling of the system during the first 1 month or so in the orbit.


Based on the specifications of the ASTRO-H cooling system, the
required power for the mechanical coolers is about 230 W, including
the nominal margin of 30\%. The total payload power, including ADR
control and signal processing system, is about 380 W\@. To cope with
this power generation, baseline design of solar paddles consists of
four panels on each side as shown in Fig.\ {spacecraft}. The mass of
the payload is about 323 kg, and the total spacecraft mass will be 615
kg which is within the capability of the current model of the Epsilon
rocket. Since we can expect some increase in the launch capability of
the rocket by the 4th mission, which is the one we hope to launch DIOS
in 2020, we may be able to consider a small upgrading regarding the
size and spacecraft mass.

\section{Science from DIOS}

For WHIM observations, realistic simulations have been carried out by
Takei et al.\cite{takei11}\ assuming instrument parameters for Xenia
which has about 4 times larger collecting area than DIOS\@. Since 
contribution of non-X-ray background is expected be negligible, we may scale the results by Takei et al.\ by increasing the observing time. The simulation included the foreground emission and required
simultaneous detection of both OVII and OVIII emission lines 
above $5\sigma$. Detection of the two lines will allow us to
estimate the average temperature of the emitting clouds. Even though
the clouds are likely to be inhomogeneous with a range of temperature,
our simulation shows that the derived temperature by assuming a
thermal equilibrium is within 30\% of the actual average temperature
of the emitting region wighted by the OVII line intensity.

\begin{figure}[!htb] \begin{minipage}{8cm}
\centerline{\includegraphics[width=0.95\textwidth,bb=0 0 388 391]{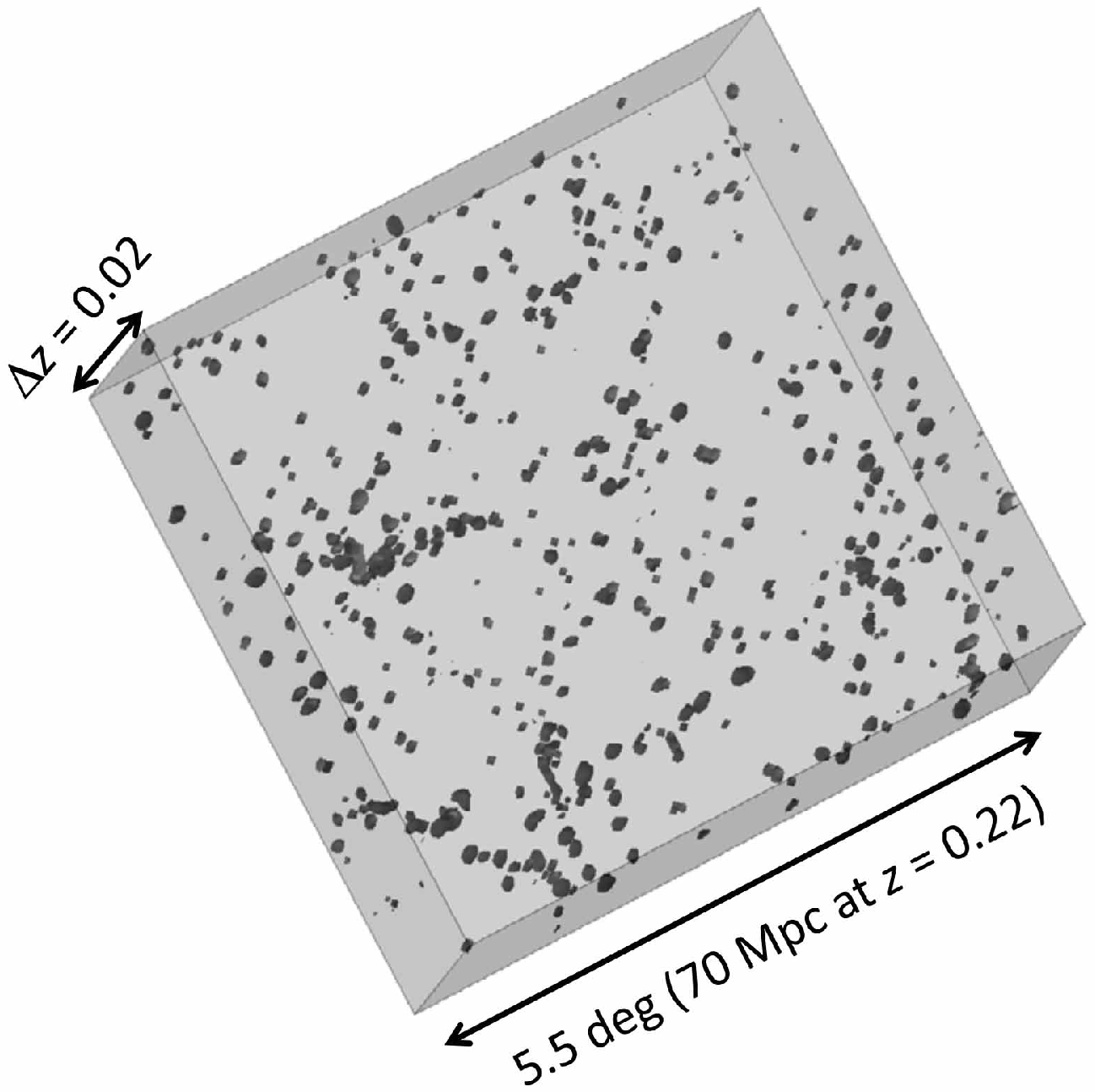}}
\caption{Expected map of WHIM based on simultaneous detection of OVII
  and OVIII lines above $5\sigma$ significance for a $5^\circ \times
  5^\circ$ sky, also sliced in the redshift space.} \label{WHIMmap}
\end{minipage} \hfill
\begin{minipage}{8cm}
\centerline{\includegraphics[width=0.95\textwidth,bb=0 0 514 435]{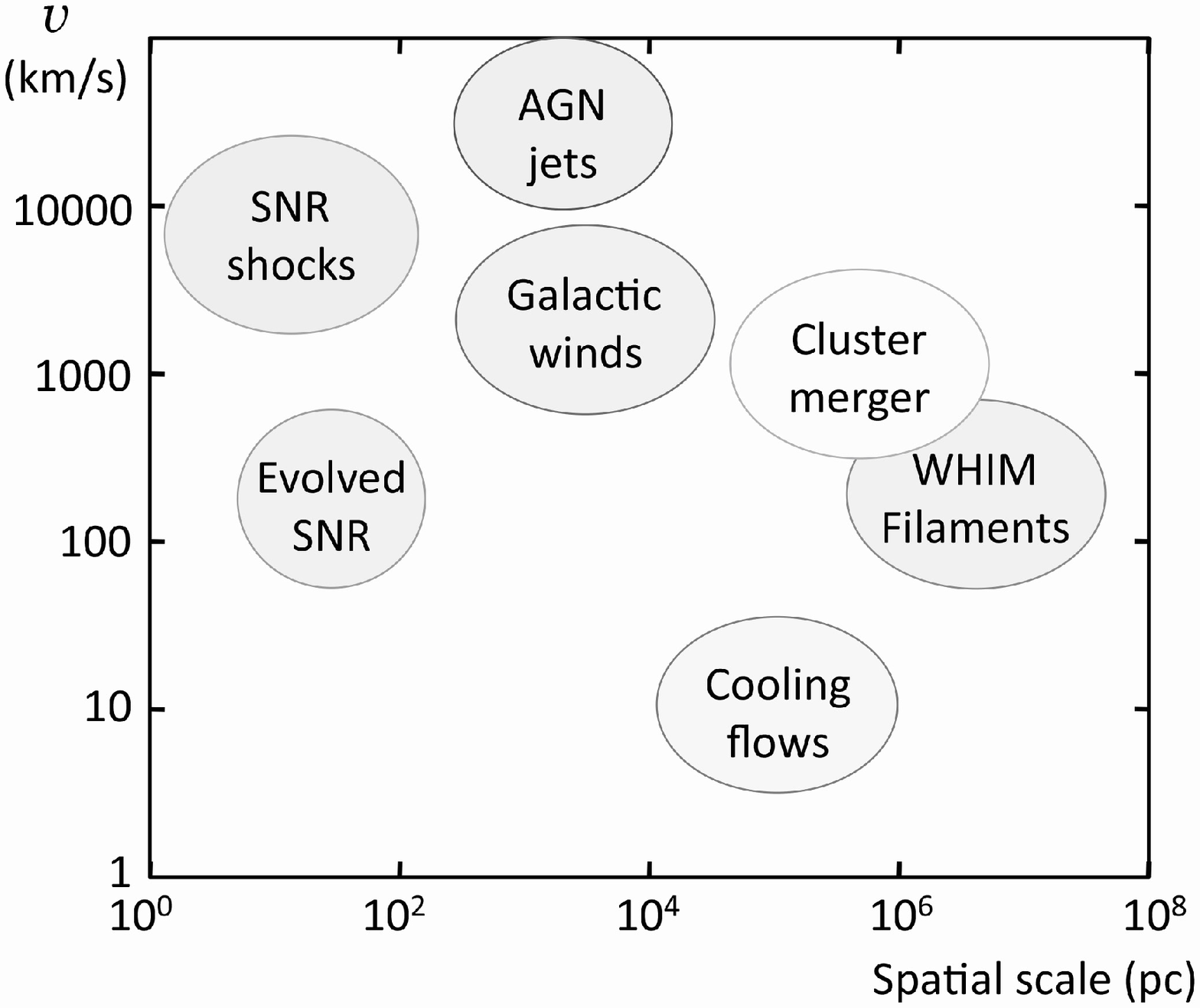}}
\caption{Large-scale gas motion expected in various systems, which will be the subject of high-resolution spectroscopy from DIOS\@.} \label{gas_dynamics}
\end{minipage}
\end{figure}

We will use energy ranges which are relatively free from foreground
emission lines, and look for WHIM oxygen lines in the corresponding
redshift intervals.  The detected WHIM clouds from our simulation show
many spatial concentrations and their distribution indicates the
presence of large-scale filaments in several
regions\cite{takei11}. Multi-pointing observations, such as $5\times
5$ degree sky surveyed in 2 years, will allow us to stack a number of
these images from different redshifts, thus yielding 3-dimensional
distribution of WHIM\@. Fig.\ \ref{WHIMmap} shows an expected map of
WHIM for a sky region of $5.5^\circ\times 5.5^\circ$ for a redshift
interval $\Delta z = 0.02$, and filamentary structures can be
recognized.

Because of the wide field of view, combined with excellent energy resolution and low intrinsic background, DIOS can perform unique
observations of extended X-ray objects which have very low surface
brightness. Here, we list several examples where DIOS is expected to
give new sciences.
\begin{enumerate}
\item Earth's magnetosphere: X-ray lines are emitted from solar-wind
  charge exchange process in magnetosheath and cusps in the Earth's
  neighborhood and DIOS will give us the structure of these emission
  regions. Clear separation of forbidden lines will be the key feature.
\item Galactic interstellar medium: hot interstellar gas is expected to
  exhibit a Galactic fountain consisting of outgoing and infalling
  gases, and their dynamical motion can be studied with DIOS.
\item Supernova remnants: Distribution and motion of all the metals in
  SNRs will be mapped and velocity structure around shocks will be
  observed.
\item Starburst galaxies: Velocity and temperature of metals-rich gas
  outflowing into the halo region of bright galaxies will be examined
  in a spatially resolved manner.
\item Radio lobes: Jets are decelerated and thermalized in radio
  lobes, and conversion process of non-thermal energy to thermal gas
  will be directly studied.
\item Clusters of galaxies: Merger shocks associated with radio
  relics, and unvirialized gas, possibly infalling, in outer regions
  of clusters will be observed, and thermalization processes will be
  closely examined.
\end{enumerate}

Fig.\ \ref{gas_dynamics} shows targets for which study of gas dynamics
in large-scale plasmas will be important. Note that most of these
objects are much brighter than WHIM and their observations will take
relatively shorter times. We will organize the observational program
of DIOS to yield the most unique and important results in an efficient
way.

\section{Possible improvement}

In Autumn 2013, JAXA has released a roadmap of Japanese program for
space science and exploration, in which 3 mission classes are
defined. The 1st category is a large mission led by ISAS, such as
ASTRO-H, within a cost of 300 Myen, the 2nd is a class of Epsilon
rocket launch with a cost of around 100 Myen, and the 3rd one is a
mission of opportunity type with about 10 Myen per year. This allows
us to make DIOS, in the 2nd category, somewhat larger than what has
been considered previously. We started looking into the following
possibilities.
\begin{enumerate}
\item Larger X-ray telescope: we may be able to extend the focal
  length from 70 cm to 1.2 m, which makes the instrument very similar
  to the one proposed for the EDGE project\cite{piro09}. This will
  give the effective area close to 1000 cm$^2$, keeping the field of
  view $0.7^\circ \times 0.7^\circ$. This option will significantly
  enhance the sensitivity to both point and diffuse sources.
\item Fast re-pointing capability: larger X-ray telescope will make
  observation of X-ray afterglow of gamma-ray bursts highly
  productive. Absorption lines from host galaxies at $z \gtrsim 3$
  would be important in exploring the epoch of galaxy formation. Small
  gamma-ray burst monitor and intelligent attitude control system will
  be considered as additional features.
\end{enumerate}
The introduction of the gamma-ray burst science, as well as the
high-resolution spectroscopy of diffuse objects, will make DIOS a very
powerful mission after ASTRO-H\@. It will also act as a pathfinder to
Athena regarding technology development of cryogen free cooling
system and TES array application including its read-out technique.

\section{Status and prospects}

In 2012, Japanese high-energy astrophysics association carried out a
review of 6 proposed missions, and evaluated DIOS to be ``S'' rank
along with PolariS (polarization measurement mission). DIOS was then
recommended to the astronomy and astrophysics sub-division committee
of the Science Council of Japan in 2013. DIOS is also listed in the
Master Plan of Large Research Projects 2014 issued by the Science
Council of Japan in March 2014, as an only one future X-ray
program. This makes DIOS recognized in the wide science community in
Japan.

The DIOS team has joined previous mission proposals aiming for a
search of dark baryons with microcalorimeter measurement of emission
lines. The proposed missions are EDGE (Cosmic Vision in 2007), Xenia
(Decadal Survey in 2009) and ORIGIN (Cosmic Vision in 2010), and the
proposing team consists of European, US and Japanese scientists. DIOS
will, thus, have international support from active groups developing
TES calorimeters and X-ray optics. International collaboration will be
an essential factor to realize the mission in view of raising the
technology readiness, enhancing the science and sharing the emission
cost.

We plan to propose DIOS to the 4th Epsilon mission with a call
expected to take place in 2015--2016 and a launch in 2019-2020, if
successfully selected. This would give us a nice extension of the
high-resolution spectroscopy science started with ASTRO-H (launch is
expected in late 2015), and technology demonstration for Athena which
will be put into orbit in 2028.

\bibliography{ohashi_spie14}   
\bibliographystyle{spiebib}   
\end{document}